\documentclass[aps,12pt,tightenlines,superscriptaddress,longbibliography,notitlepage]{revtex4-1}

\usepackage{graphicx,mdwlist,color,wrapfig,revsymb}

\begin{document}

\title{Observation of the fractional a.c. Josephson effect and the signature of Majorana particles}

%Observation of fractional ac Josephson effect: the signature of Majorana particles}

\author{Leonid P. Rokhinson}
\thanks{To whom correspondence should be addressed. E-mail: leonid@purdue.edu}
\affiliation{Department of Physics, Purdue University, West Lafayette, Indiana
47907 USA} \affiliation{Birck Nanotechnology Center, Purdue University, West
Lafayette, Indiana 47907 USA}

\author{Xinyu Liu}
\author{Jacek K. Furdyna}
\affiliation{Department of Physics, University of Notre Dame, Notre Dame,
Indiana 46556 USA}

%\date{Submitted to Science on March 20, 2012}
\date{\today, ---= final draft: \jobname.tex =---}
%\pacs{+75.50.Pp, 73.20.Fz, 71.23.-k}

\begin{abstract}
Topological superconductors which support Majorana fermions are thought to be
realized in one-dimensional semiconducting wires coupled to a superconductor
\cite{Lutchyn2010a,Oreg2010,Alicea2011}. Such excitations are expected to
exhibit non-Abelian statistics and can be used to realize quantum gates that
are topologically protected from local sources of decoherence
\cite{Kitaev2001,Kitaev2003}. Here we report the observation of the fractional
a.c. Josephson effect in a hybrid semiconductor/superconductor InSb/Nb nanowire
junction, a hallmark of topological matter. When the junction is irradiated
with a radio-frequency $f_0$ in the absence of an external magnetic field,
quantized voltage steps (Shapiro steps) with a height $\Delta V=hf_0/2e$ are
observed, as is expected for conventional superconductor junctions, where the
supercurrent is carried by charge-$2e$ Cooper pairs. At high magnetic fields
the height of the first Shapiro step is doubled to $hf_0/e$, suggesting that
the supercurrent is carried by charge-$e$ quasiparticles. This is a unique
signature of Majorana fermions, elusive particles predicted \textit{ca.} 80
years ago \cite{Majorana37}.
\end{abstract}

\maketitle

In 1928 Dirac reconciled quantum mechanics and special relativity in a set of
coupled equations, which became the cornerstone of quantum
mechanics\cite{Dirac1928}. Its main prediction that every elementary particle
has a complex conjugate counterpart -- an antiparticle -- has been confirmed by
numerous experiments. A decade later Majorana showed that Dirac's equation for
spin-1/2 particles can be modified to permit real
wavefunctions\cite{Majorana37,Wilczek2009}. The complex conjugate of a real
number is the number itself, which means that such particles are their own
antiparticles. While the search for Majorana fermions among elementary
particles is ongoing \cite{Cho2011}, excitations with similar properties may
emerge in electronic systems\cite{Kitaev2001}, and are predicted to be present
in some unconventional states of matter
\cite{moore91,Sengupta2001,DasSarma2006,Read2000,fu08,sau10}.

Ordinary spin-1/2 particles or excitations carry a charge, and thus cannot be
their own antiparticles. In a superconductor, however, free charges are
screened, and charge-less spin-1/2 excitations become possible. The BCS theory
allows fermionic excitations which are a mixture of electron and hole creation
operators, $\gamma_i=c_i^{\dag}+c_i$. This creation operator is invariant with
respect to charge conjugation, $c_i^{\dag}\leftrightarrow c_i$. If the energy
of an excitation created in this way is zero, the excitation will be a Majorana
particle. However, such zero-energy modes are not permitted in ordinary
$s$-wave superconductors.

The current work is inspired by the paper of Sau {\it et al.} \cite{sau10} who
predicted that Majorana fermions can be formed in a coupled
semiconductor/superconductor system. Superconductivity can be induced in a
semiconductor material by the proximity effect. At zero magnetic field
electronic states are doubly-degenerate and Majorana modes are not supported.
In semiconductors with strong spin-orbit (SO) interactions the two spin
branches are separated in momentum space, but SO interactions do not lift the
Kramer's degeneracy. However, in a magnetic field
$\mathbf{B}\bot\mathbf{B_{so}}$ there is a range of energies where double
degeneracy is lifted\cite{Quay2010}, see schematic in Fig~\ref{f-dev}c. If the
Fermi energy $E_F$ is tuned to be within this single-mode range of energies,
$E_Z>\sqrt{\Delta^2+E_F^2}$, (where $\Delta$ is the proximity gap,
$E_Z=g\mu_BB/2$ is the Zeeman energy, $\mu_B$ is the Bohr magneton, and $g$ is
the Land\'e $g$-factor), the proximity effect from a conventional $s$-wave
superconductor induces $p$-wave pairing in the semiconductor material and
drives the system into a topological superconducting state which supports
Majorana particles. Theoretically, it has been predicted that proper conditions
for this to occur can be realized in 2D \cite{sau10,Alicea2010} and, most
relevant to the current work, in 1D systems\cite{Lutchyn2010a,Oreg2010}.
Moreover, multiband nanowires should also support topological
superconductivity\cite{Lutchyn2010,Potter2010,Stanescu2011}.

What are the experimental signatures of Majorana particles? Majorana particles
come in pairs, and zero energy Andreev end-modes localized at the ends of a
wire can be probed in tunneling
experiments\cite{Sengupta2001,Law2009,sau2010a}. Indeed, there are reports of
zero bias anomaly observed in topological
insulator/superconductor\cite{Koren2011} and
semiconductor/superconductor\cite{Mourik2012,Deng2012} structures. However,
conductivity enhancement near zero bias can also be a signature of diverse
phenomena in mesoscopic physics, such as the Kondo effect in quantum
dots\cite{goldhaber98a,rokhinson99a} or the ``0.7 anomaly'' in
nanowires\cite{cronenwett02,rokhinson06}. Fusion of two Majorana modes produces
an ordinary fermion and, uniquely to Majorana particles, modifies periodicity
of the Josephson relation from $2\pi$ (Cooper pairs) to $4\pi$ (Majorana
particles)\cite{Kitaev2001,Kwon2004,Fu2009,Lutchyn2010a,Akhmerov2011}. In the
dc Josephson effect, fluctuations between filled and empty Majorana modes will
mask the $4\pi$ periodicity and, indeed, we observe only $2\pi$ periodicity in
a dc SQUID configuration. In the ac Josephson effect, however, the $4\pi$
periodicity due to Majorana modes should be fully revealed.

\begin{figure}
\def\ffile{f-dev}
\includegraphics[width=0.5\textwidth]{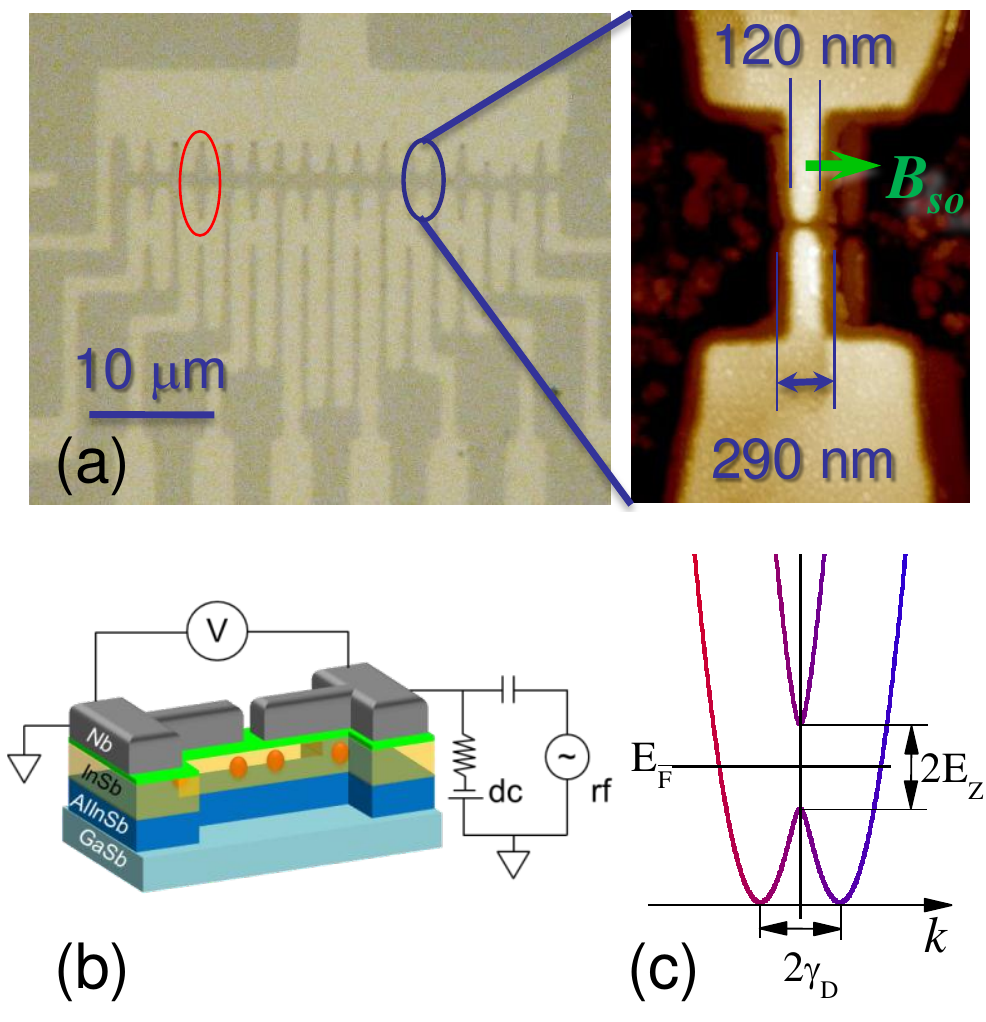}
\caption{\textbf{Devices layout.} (a) Optical image of a sample with several
devices. A single dc SQUID device is outlined with a red oval. On the enlarged
AFM image a single Josephson junction is shown. The light areas are Nb. A light
brown halo around Nb is a thin 2-3 nm Nb layer which defines the width of the
semiconductor wire after wet etching. The direction of the spin-orbit field
$B_{so}$ is indicated by the green arrow. (b) A schematic view of the device,
the dots mark the expected positions of Majorana particles inside the InSb
nanowire. (c) Energy dispersion in a material with spin-orbit interaction in
the presence of magnetic field $B\bot B_{SO}$.}
\label{\ffile}
\end{figure}

In our experiments Nb/InSb/Nb Josephson junctions (JJs) are fabricated
lithographically from a shallow InSb quantum well. Superconductivity in InSb is
induced by the proximity effect from a Nb film placed on top of the InSb
nanowire. The self-aligning fabrication process which we use is described in
the Supplementary Information. A pattern of multiple JJs is defined by e-beam
lithography, and a 45 nm layer of Nb is deposited by dc sputtering on top of
the InSb quantum well. An image of a JJ region is shown in Fig.~\ref{f-dev}a. A
weak link is formed between two 120 nm-wide and 0.6 $\mu$m-long Nb wires, with
gaps ranging from 20 to 120 nm in different devices. During deposition of Nb, a
thin (2-4 nm) layer of Nb is extended 70-80 nm outside the pattern, including
the space inside the gaps (a brown halo around the Nb wire on the AFM image).
This layer is used as an etch mask to define a nanowire in the underlying
semiconductor self-aligned to the Nb wire. After the etching a continuous
$w\lesssim 290$ nm-wide InSb wire is formed under the Nb, as shown
schematically in Fig.~\ref{f-dev}b. The thin Nb layer is not conducting at low
temperatures, so that the supercurrent is carried by the proximity-induced
superconductivity in InSb.

The InSb wires have rectangular cross section $w\gg d$ and SO interactions are
dominated by the Dresselhaus term $H_D=\gamma_D\langle
k_z^2\rangle(k_x\sigma_x-k_y\sigma_y)$, where $\langle k_z^2\rangle=(\pi/d)^2$,
$d=20$ nm is the quantum well thickness, $\gamma_D=760$ eV$\cdot$\AA$^3$ for
InSb, $\sigma_i$ are Pauli matrices, and $\hat{x}$ and $\hat{y}$ are the
principal crystallographic axes. We estimate $E_{SO}\approx1$ meV\cite{sup}.
The wires are oriented along the [110] crystallographic direction, and expected
direction of the effective spin-orbit magnetic field $\mathbf{B_{so}}$ is
perpendicular to the current, as indicated by the green arrow on the AFM image.
At high fields ($E_Z>\sqrt{\Delta^2+E_F^2}$) Majorana particles are expected to
form inside the InSb wire close to the ends of the Nb wires. At these fields
the supercurrent is dominated by the fusion of two Majorana particles across
the gap, which amounts to the $1e$ charge transfer. From the lithographical
dimensions we estimate that only a few (1-3) one-dimensional subbands should be
populated in InSb nanowires, however, we expect the density of states in the
nanowires to be modified by the strong coupling to Nb and the actual number of
filled subbands may be larger.

\begin{figure}
\def\ffile{f-iv}
\includegraphics[width=0.7\textwidth]{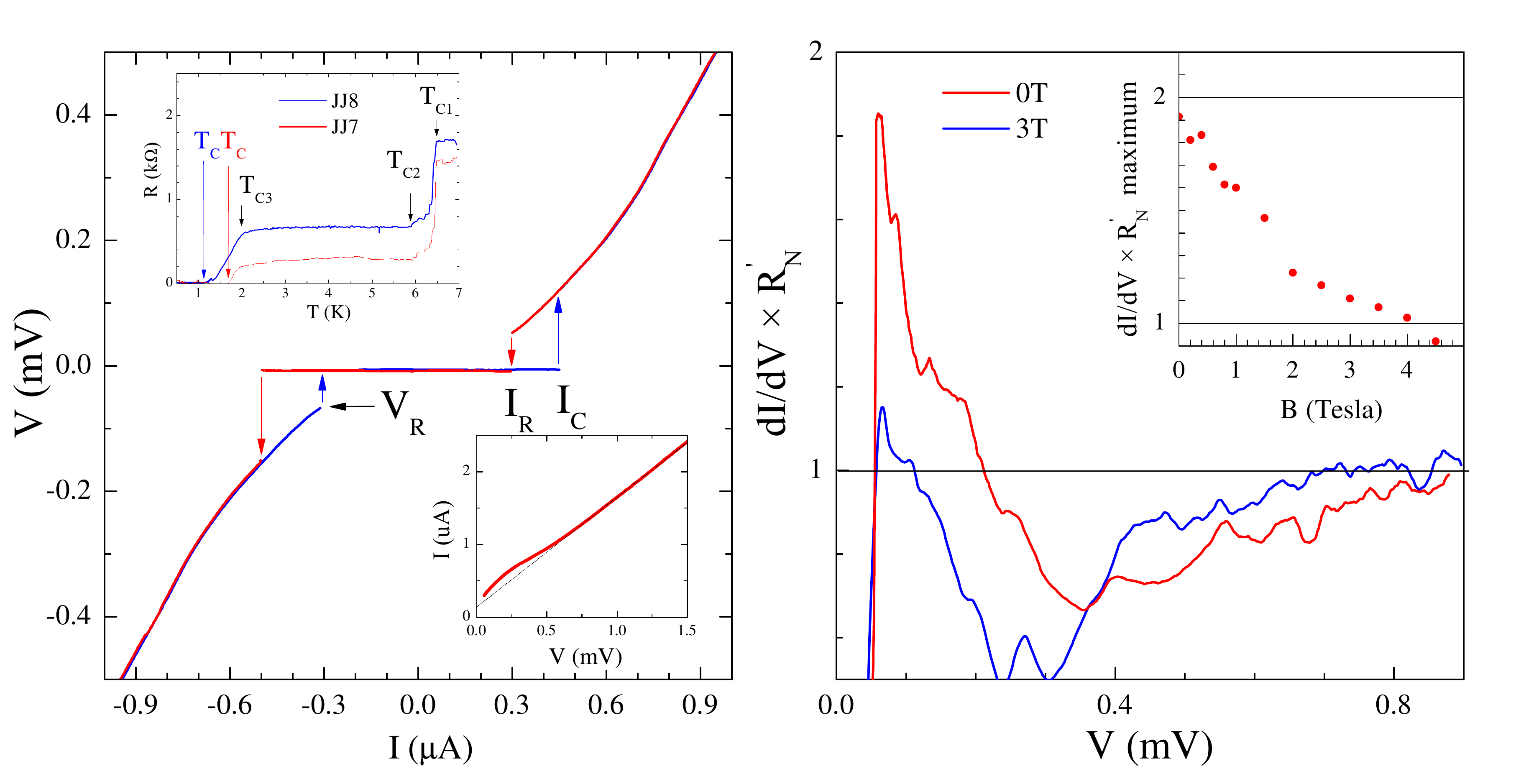}
\caption{\textbf{Characterization of Josephson junctions.} Left: $V(I)$
characteristic of a Nb/InSb/Nb Josephson junction JJ8 (40 nm gap) measured at
20 mK. In the bottom inset the same characteristic is plotted as $I$ vs $V$,
where the black line is an extrapolation from high $V$.  At $V<0.5$ mV the
excess current is clearly seen. The top inset shows the temperature dependence
for JJ7 (30 nm gap) and JJ8. Right: Normalized differential conductance is
plotted as a function of voltage across JJ8, with $R_N^{'}=650\ \Omega$. The
enhancement of the current seen at low $V$ is a signature of Andreev
reflection. The maximum enhancement is plotted as a function of $B$ in the
inset.}
\label{\ffile}
\end{figure}

As devices are cooled down, a series of superconducting transitions
$T_{c1}-T_{c3}$ and $T_c$ are observed (Fig.~\ref{f-iv}). The first transition,
$T_{c1}\sim 6.4$ K, is for wide areas, $T_{c2}=5.8$ K and $T_{c3}=1.9$ K are
for the 1 and 0.12 $\mu$m-wide Nb wires, and $T_{c}$ is for the JJs. From
$T_c=1.17$ K for JJ8 we estimate a proximity gap $\Delta\approx 180\ \mu$eV and
a semiconductor-superconducting coupling $\lambda\approx2.6\Delta$
\cite{Sau2012}. Lithographically our devices consist of two JJs in parallel,
and we can measure the ratio of the critical currents in the two arms
$r=I_{c1}/I_{c2}$ by measuring current modulation in a dc SQUID configuration.
The ratio $r=7.3$ for JJ7 and $r>10$ for JJ8, indicating that conduction is
dominated by a single junction. In the following analysis we will treat our
devices as containing a single JJ.

\begin{figure*}
\def\ffile{f-rf}
\includegraphics[width=0.95\textwidth]{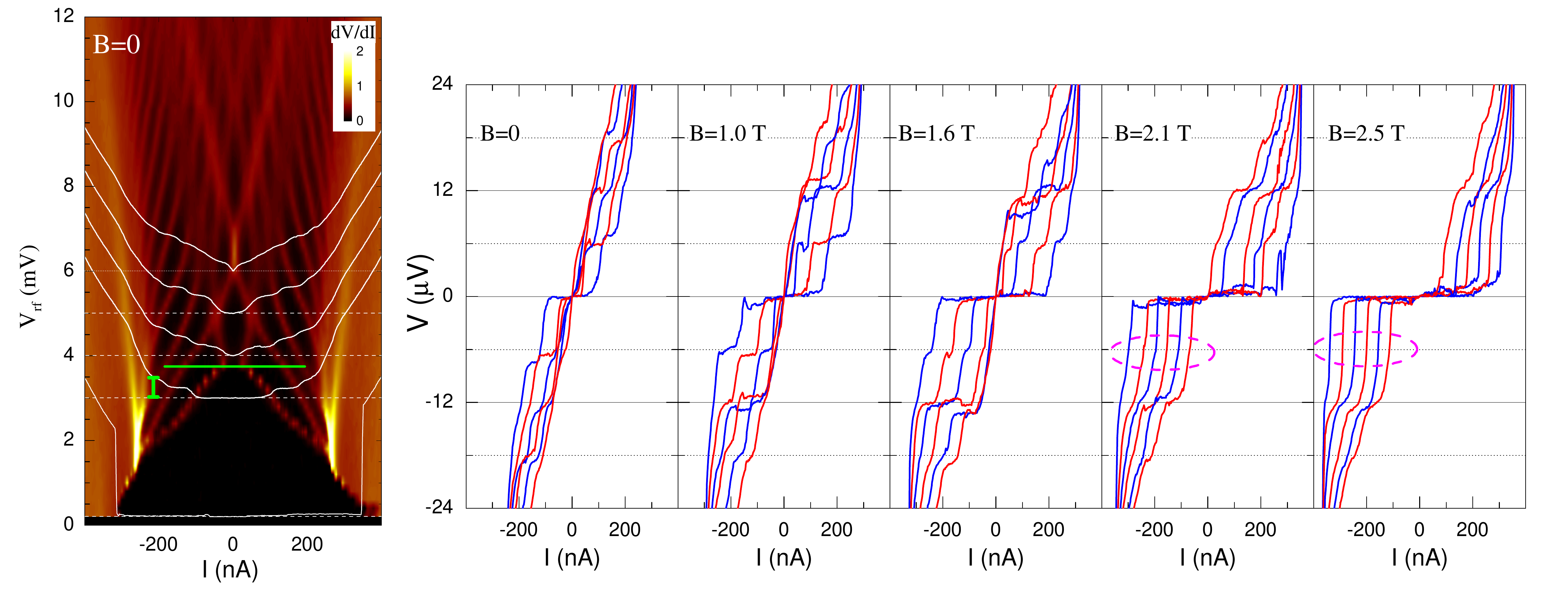}
\caption{\textbf{ac Josephson effect and Shapiro steps.} Left: differential
resistance $dV/dI$ (in k$\Omega$) of JJ8 is plotted as a function of the rf
amplitude $V_{rf}$ and dc current $I$ for $f_0=3$ GHz. The data is measured
with low frequency (17 Hz) ac excitation $I_{ac}=2$ nA at $T=20$ mK and $B=0$.
$|V(I)|$ characteristics at $V_{rf}=0.2$, 3, 4, 5 \& 6 mV are shown as white
lines; their zero is shifted vertically and is marked by dashed white lines.
The small vertical green bar indicates the scale of 12 $\mu$V. Right: $V(I)$
characteristics of the JJ8 in the presence of $\mathbf{B}\|\mathbf{I}$ measured
with $V_{rf}$ between 3 and 6 mV in 0.6 mV increments. For $B<2$ T, Shapiro
steps with a height $\Delta V=h\nu_{rf}/2e=6$ $\mu$V are clearly observed. For
$B>2$ T the plateau at 6 $\mu$V disappears, and the first step is observed at
12 $\mu$V. This doubling of the first Shapiro step is a signature of the ac
fractional Josephson effect, and is a hallmark of a topological
superconductivity.}
\label{\ffile}
\end{figure*}

As seen in Fig.~\ref{f-iv}, $V(I)$ characteristic for JJ8 measured at the base
temperature of 20 mK exhibits a clear supercurrent region ($V=0$), with an
abrupt appearance of a finite voltage. A small hysteresis is observed for the
return critical current $I_R\sim I_c$, characteristic of a resistively shunted
JJ in an intermediate damping regime\cite{Tinkham-book96}. Indeed, we measure
high leakage to the substrate, and estimate the shunting resistance
$R_s\lesssim 1$ k$\Omega$. The measured resistance in the normal state is $R_N^{'}=650\
\Omega$, and the actual normal resistance of the JJ8 $R_N\approx2-6$ k$\Omega$,
consistent with the number of 1D channels estimated from the size of the wire.
The product $R_N I_c\approx 1$ mV$>2\Delta$ indicates that JJ8 is in a clean
limit (weak link length $L_{eff}<\xi,l$, where $\xi$ is the coherence length
and $l$ is the mean free path), a proper condition for the formation of
Majorana particles.

Normalized differential conductance, plotted in the right panel of
Fig.~\ref{f-iv}, shows enhancement at low voltages. This excess current is a
signature of Andreev reflection\cite{Andreev1964,Doh2005}. Most important for
our measurements is that the excess current, and thus coherent electron
transport, is observed at high in-plane magnetic fields up to 4 T, as shown in
the the inset.

In the presence of a microwave excitation, phase locking between the rf field
and the Josephson supercurrent gives rise to constant-voltage Shapiro steps in
the $V(I)$ characteristics at $V_n=nhf_0/q$, where $h$ is Planck's constant,
$q$ is the charge of quasiparticles, $f_0$ is the microwave frequency, and
$n=0,\pm1,\pm2...$ \cite{Shapiro1963}. Shapiro steps for $f_0=3$ GHz are shown
in Fig.~\ref{f-rf}. At $B=0$ we observe steps with the height $\Delta V=6\
\mu$V, consistent with the Cooper pair tunneling ($q=2e$). $\Delta V$ scales
linearly with $f_0$ \cite{sup}. The evolution of the steps with $V_{rf}$ is
best visualized in the $dV/dI$ plots, where steps with $0<n<10$ are seen at
high rf powers. A transition from low to high rf power regime is clearly seen
in the $dV/dI$ plot near $V_{rf}\approx 4$ mV and is marked by a green
horizontal line. At high rf powers $V_{rf}>4$ mV the evolution of the width of
the Shapiro steps $\Delta I_n$ follows a Bessel function pattern as a function
of power, $\Delta I_n=A|J_n(2e v_{rf}/hf_0)|$, where $v_{rf}$ is the rf
amplitude at the junction. We can find the rf power attenuation from the fit to
the $\Delta I_0$, $v_{rf}=5\cdot 10^{-3} V_{rf}$ for $f_0=3$ GHz,
Fig.~\ref{f-dI}. Here $V_{rf}$ is the rf amplitude at the top of the fridge.
Thus, $V_{rf}=4$ mV corresponds to $v_{rf}=20\ \mu$V$\ \approx V_R/2\sqrt{2}$,
where $V_R\approx60\ \mu$V, see Fig.~\ref{f-iv}.

For $V_{rf}<4$ mV the junction is in the small microwave signal regime
\cite{LikharevJJbook}. The linear response of a JJ has a singularity at
$\omega=\pm\omega_J$, where $\Omega_J=2e/\hbar V$ is the Josephson frequency,
and the JJ performs a parametric conversion of the external frequency. The JJ8
is in the intermediate damping regime and the $V(I)$ characteristic is expected
to become non-hysteretic in the vicinity of the first step. Indeed, we observe
no hysteresis for $V_{rf}>1.8$ mV. While nonlinear effects can be present at
high $I$ and $V_{rf}$, we want to stress that the first step at the onset of
the normal state is due to phase locking between the external and the Josephson
frequencies, $\omega=\pm\omega_J$.

When in-plane magnetic field $\mathbf{B}\| \mathbf{I}$ is applied, Shapiro
steps at $V=6, 12$ and 18 $\mu$V are clearly visible at low fields, $B<2$ T.
Steps at 12 and 18 $\mu$V remain visible up to $B\approx 3$ T, while the step
at 6 $\mu$V disappears above $B\approx2$ T. The disappearance of all steps
above 3 T is consistent with suppression of the excess current and Andreev
reflection at high fields.

\begin{figure}
\def\ffile{f-dI}
\includegraphics[width=0.7\textwidth]{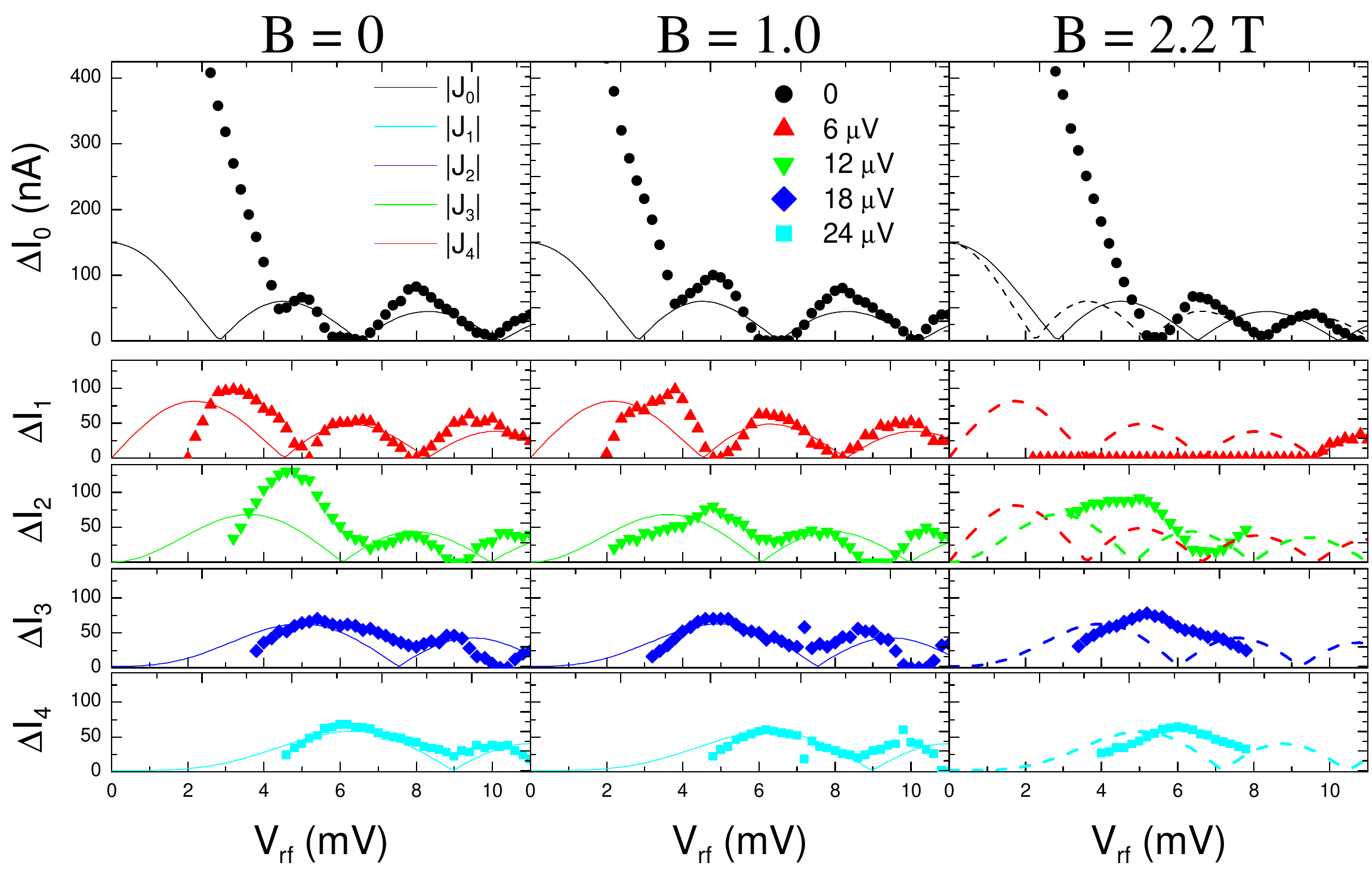}
\caption{\textbf{Evolution of Shapiro steps with rf power.} The width of the
first five Shapiro steps are plotted as a function of the rf field amplitude
$V_{rf}$. Lines are Bessel functions $A|J_n(\beta V_{rf})|$, with $\beta=0.84$
mV$^{-1}$ (solid) and $\beta=1.04$ mV$^{-1}$ (dashed).}
\label{\ffile}
\end{figure}

Quantitative comparison of the width of the Shapiro steps $\Delta I_n(V_{rf})$
for different $B$ extracted from $dV/dI$ data is plotted in Fig.~\ref{f-dI}.
For $B<2$ T steps at 0, 6, 12, 18 and 24 $\mu$V evolve according to Bessel
functions with the amplitude $A\approx 150$ nA. For $B>2$ T the step at 6
$\mu$V vanishes at low $V_{rf}$, and re-appears at high $V_{rf}>10$ mV for
$B=2.2$ T. We also observe that the low field rf attenuation does not fit the
evolution of the $\Delta I_0$ plateau. Moreover, a minimum of the 12 $\mu$V
plateau at $V_{rf}\approx7$ meV coincides with the minimum of the $|J_1|$
Bessel function, suggesting that indeed the 12 $\mu$V plateau became the $n=1$
Shapiro step. Evolution of plateaus at $B>2$ T is poorly described by Bessel
functions, suggesting that oscillations with different frequencies may
contribute to the width if the plateaus. We emphasize that the doubling of the
Shapiro step height is a unique signature of a topological quantum phase
transition.

Theoretically it has been argued that Josephson current with both $2\pi$
($I_c\sin(\phi)$) and $4\pi$ ($I_M\sin(\phi/2)$) periodicity should be present
in the topological state\cite{Alicea2011,Jiang2011,Pikulin2011,San-Jose2011a},
especially in a multichannel wires. However, in current--biased junctions odd
steps are expected to vanish even in the presence of large supercurrents
carried by the charge-$2e$ quasiparticles $I_c\gg I_M$ \cite{Dominguez2012}. In
this case the width of even steps is expected to be defined primarily by $I_c$
and the coefficient $A$ remains almost unchanged across the transition, as is
observed experimentally, especially if a large number of subbands is occupied
in the nanowire due to the strong coupling to the superconductor. At high
voltages across the junction we expect enhanced mixing between gapless and
gapped modes, which may explain the prominence of the 18 $\mu$V and higher
plateaus at $B>2$ T.

\textbf{Acknowledgements} The work was partially supported by ARO grant
W911NF-09-1-0498 (L.P.R.) and by NSF grant DMR10-05851 (J.K.F., X.L.). L.P.R.
benefited from discussions with Roman Lutchyn.

\textbf{Authors contributions} L.P.R. conceived and performed the experiments;
J.K.F. and X.L. designed and grew the heterostructures; all authors contributed
to the writing of the manuscript.

%\bibliographystyle{naturemag}
%\bibliography{rohi,majorana}

\clearpage

\renewcommand{\thefigure}{S\arabic{figure}}
\renewcommand{\theequation}{S\arabic{equation}}

\begin{center}
\textbf{\Large Supplementary Information} \\
\vspace{0.2in} \textsc{Observation of fractional ac Josephson effect: the signature of Majorana particles}\\
{\it Leonid P. Rokhinson, Xinyu Liu and Jacek K. Furdyna}
\end{center}

\tableofcontents

\section{Analysis of the parameter space to observe fractional Josephson effect}
\label{samples}

In order to form Majorana fermions in a nanowire, several conditions have to be
satisfied. The most stringent is lifting of the Kramers degeneracy,
$E_Z>\sqrt{\Delta^2+E_F^2}$, where $E_Z$ is Zeeman energy $E_Z=g^*\mu_BB$,
$g^*=50$ for InSb, $\mu_B$ is Bohr magnetron, and $B$ is an external magnetic
field. At the same time we need the proximity gap to be larger than the
Josephson frequency, $\Delta>h\omega_J=6$ $\mu$eV for 3 GHz, or
$\Delta>2h\omega_J$ in the topological phase. The proximity gap $\Delta$
depends on the semiconductor-superconductor coupling $\lambda$, Zeeman energy
and spin-orbit coupling \cite{Sau2012}:
\begin{equation}
\Delta=\Delta_s
\frac{\lambda}{\lambda+\Delta_s}\frac{E_{SO}}{\sqrt{E_{SO}^2+E_Z^2}}.
\label{delta}
\end{equation}
The maximum $\Delta$ cannot exceed  the superconducting gap in narrow Nb wires,
$\Delta^w_s=290$ $\mu$eV for $T_c^w=2$ K, and is maximized for large coupling
$\lambda\gtrsim\Delta_s$. Large coupling, though, increases electron density in
the semiconductor which, in turn, increases $E_F$, thus requiring large fields
to lift the Kramers degeneracy.

\begin{figure}[t]
\def\ffile{fs-param}
\includegraphics[width=0.7\textwidth]{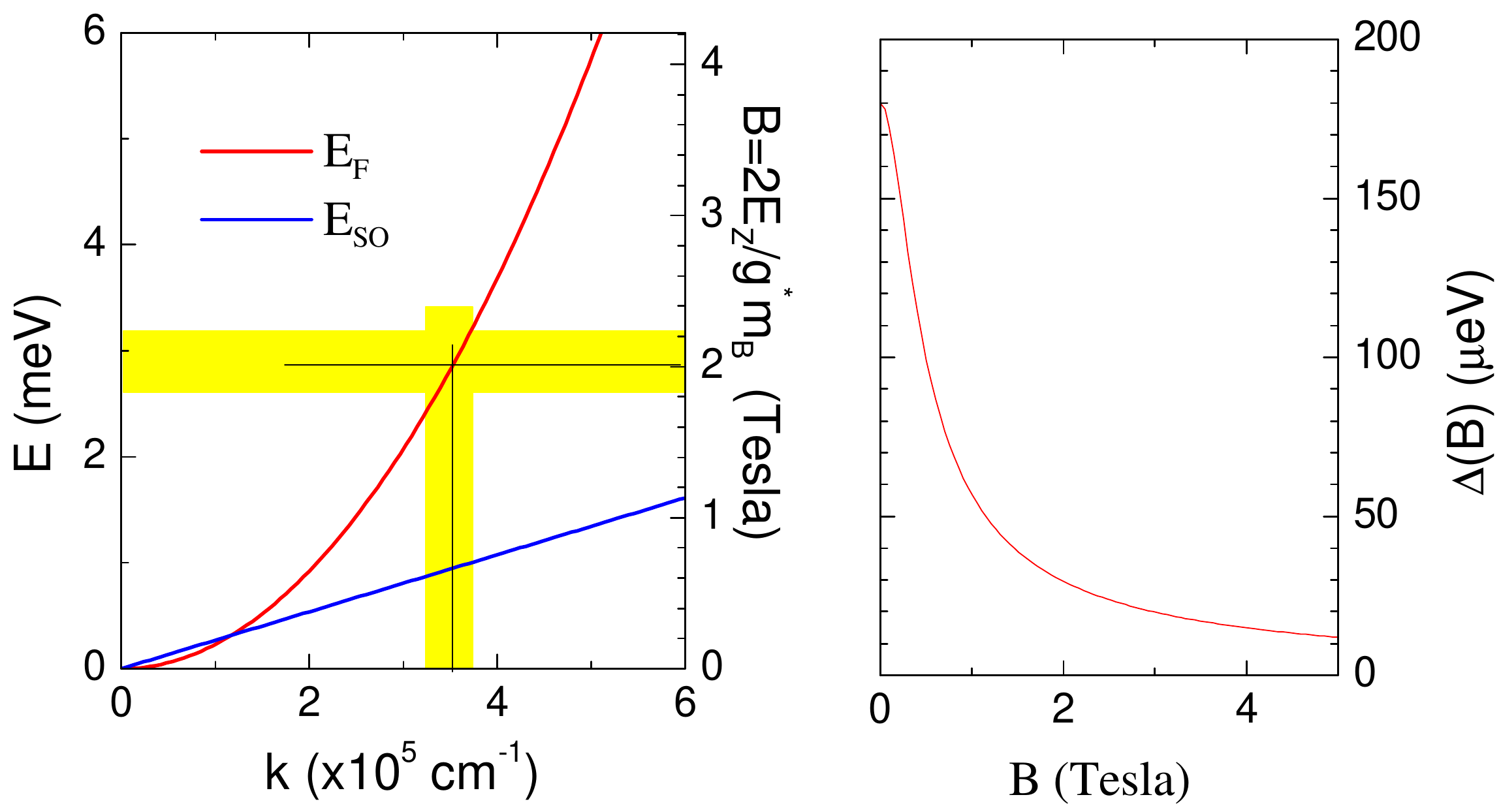}
\caption{Left: Fermi energy $E_F$ and energy of spin-orbital coupling $E_{SO}$
(for 20 nm InSb quantum well) are plotted as a function of carrier momentum
$k$. The right axis shows magnetic fields which correspond to the Zeeman energy
of the left axis. Right: expected field dependence  of the proximity gap
(Ed.~\ref{delta}) for $E_{SO}=1$ meV and $\Delta(0)=180\ \mu$eV.}
\label{\ffile}
\end{figure}

Let us analyze the parameters of the JJ8 device. We measured $T_c=1.17$ K, thus
$\Delta=1.76 k_BT/e=180$ $\mu$eV and $\lambda=470$ $\mu$eV, or $\lambda\approx
2.6\Delta$. In our sample geometry we cannot measure electron density
independently, but we can assume that at $B=2$ Tesla (observed topological
phase transition) $E_Z=E_F$. We plot $E_F= (\hbar k)^2/2m^*$, $E_z=g\mu_BB/2$
and $E_{SO}=\sqrt{2}\gamma_D\langle k^2_z\rangle k$ as a function of $k$ in
Fig.~\ref{fs-param}. Here we use $\gamma_D=760$ eV$\cdot$\AA$^3$, $\langle
k^2_z\rangle=(\pi/d)^2$, effective mass $m^*=0.015m_0$, effective $g$-factor
$g=50$, and the quantum well thickness $d=20$ nm. At $B=2$ T ($E_Z=2.8$ meV)
the condition $E_F=E_Z$ translates into $k\approx3.5\cdot10^5$ cm$^{-1}$.
Spin-orbit coupling for this $k$ is estimated $E_{SO}\approx1$ meV. The
expected field dependence of the proximity gap $\Delta(B)$ is plotted on the
right plot for $E_{SO}=1$ meV. We see that at $B=2$ T $\Delta=30$ $\mu$eV and
the condition $\Delta>2 h\omega_J$ is satisfied.

We observe Shapiro steps up to $\approx 3$ T, which is consistent with the
theoretically expected gap to be only $20$ $\mu$eV at that field.
Experimentally we measure a weaker $I_c$ vs $B$ dependence than the one
predicted by Eq.~(\ref{delta}), see Section \ref{Ic-T-dependence}.

The maximum width of the InSb wire is 290 nm, and we expect the actual InSb
wire width to be reduced due to side etching and surface depletion. For an InSb
wire with width $w=100-250$ nm the energy separation between the first 2 energy
levels $3h^2/8m^*w^2$ is 1.2-8 meV and only $<3$ subbands are expected to be
occupied for $E_F<3$ meV. We note, though, that strong coupling to Nb modifies
the density of states in the InSb wire and the actual number of filled subbands
may be larger. \textbf{Thus, we conclude that experimental parameters for the
JJ8 device satisfy the requirements for the observation of Majorana fermions.}

\section{Fabrication of Josephson junctions}

\begin{figure}
\def\ffile{fs-dev-sup}
\includegraphics[width=.9\textwidth]{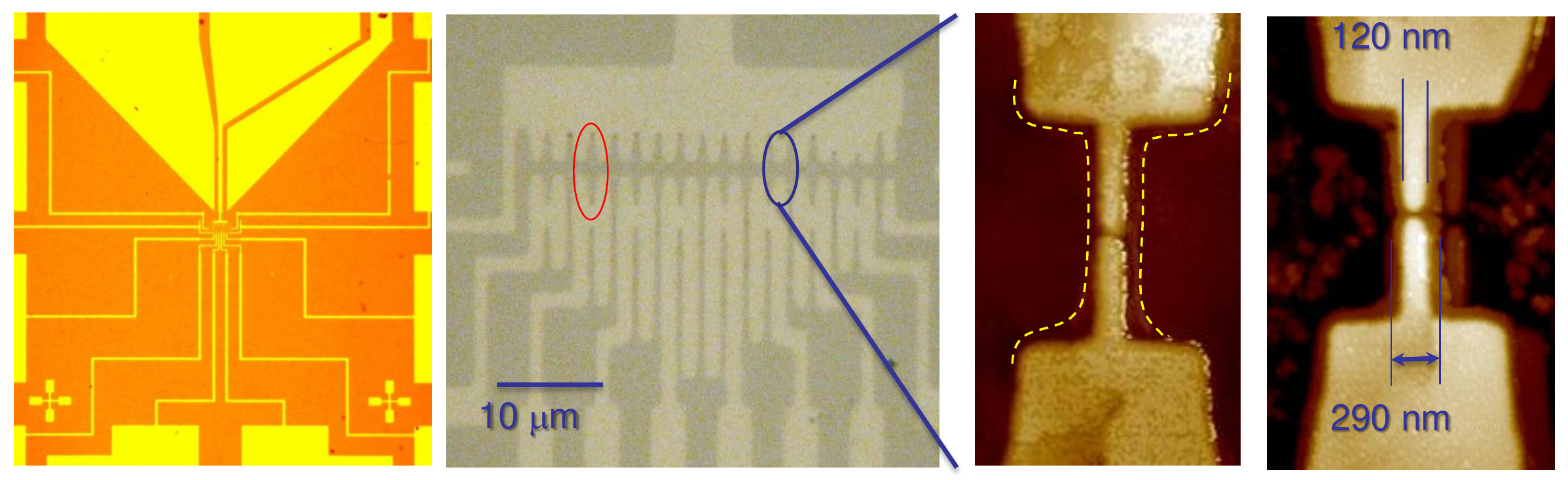}
\caption{Optical image of a device and AFM micrographs of a Josephson junction
before and after wet etching. Yellow dashed line outlines the extent of the
thin Nb layer.}
\label{\ffile}
\end{figure}

The starting material is undoped
In$_{0.6}$Ga$_{0.4}$Sb/InSb/In$_{0.6}$Ga$_{0.4}$Sb (3nm/20nm/3nm) quantum well
grown by molecular beam epitaxy on a Te-doped (001) GaSb substrate. A thick
graded In$_{x}$Ga$_{1-x}$Sb ($x=.17-.6$) buffer and a 120nm
In$_{0.77}$Al$_{0.33}$Sb barrier were grown between the wafer and the quantum
well for strain relaxation and electron confinement. A pattern of multiple JJs
is defined by e-beam lithography, and a 45 nm thick layer of Nb is deposited by
dc sputtering. Surface oxide is removed by dipping the sample in HF:DI (1:10)
for 20 seconds prior to Nb deposition. An optical image of a multi-device
sample is shown in Fig.~\ref{fs-dev-sup}. Weak links are formed between two 120
nm-wide and 0.6 $\mu$m-long Nb wires, with gaps in 20-120 nm range.

The key processing step is self alignment of the Nb and InSb wires. We use
double layer MMA/PMMA photoresist which creates an undercut after e-beam
exposure and development. During Nb sputtering at $0.1$ mTorr of Argon some Nb
is scattered into this undercut area and a thin (2-4 nm) layer of Nb is formed,
extending 70-80 nm outside the pattern, including the space inside the gaps (a
brown halo around the wire in the AFM image; see also Fig.~\ref{fs-GaAs}). This
layer is used as an etch mask to define a nanowire in the underlying
semiconductor, which becomes self-aligned to the Nb wire. Etching in
H$_2$SO$_4$:H$_2$O$_2$:H$_2$O (1:8:1000) for 30 sec removes 60 nm of
semiconductor and a continuous $\approx 170$ nm-wide InSb wire is formed under
the Nb (we expect that the width of the wire is reduced during the wet etching
step by $\sim 60$ nm from each side).

\section{Proximity effect}
\label{GaAs}

\begin{figure}
\def\ffile{fs-GaAs}
\includegraphics[width=.6\textwidth]{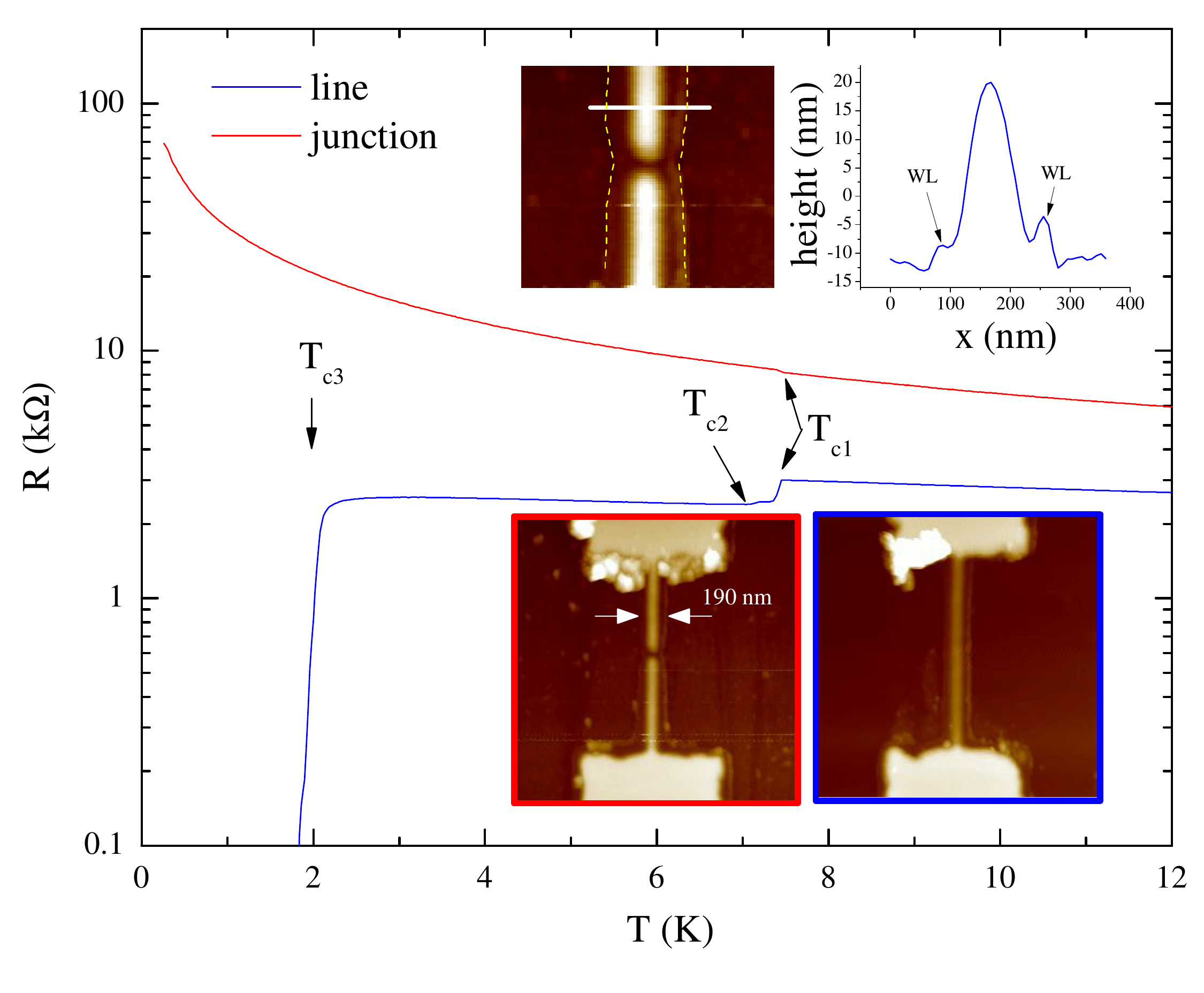}
\caption{Temperature dependence of a Nb junction and a continuous wire
fabricated on a semi-insulating GaAs substrate. The wires are fabricated
identically to the JJ on InSb substrates. In the bottom inset AFM micrographs
of the junction and the wire devices are shown ($2\ \mu\mathrm{m}\times2\ \mu$m
scan size). On the top image a zoom of the gap region is shown. A 4 nm-high
wetting layer around the wire and within the gap is clearly seen and is
outlined with a dashed yellow line. The wire profile along the white line is
shown on the right, and the thin Nb layer is marked as WL.}
\label{\ffile}
\end{figure}

In order to verify that we observe proximity-induced superconductivity, we
fabricated several ($>10$) test devices on a semi-insulating GaAs substrate
with the Nb pattern identical to the JJ devices, see inset in
Fig.~\ref{fs-GaAs}. Continuous wires show the expected superconducting phase
transitions at $T_{c1}=7.5$ K for wide regions ($>6 \mu$m), $T_{c2}=7.1$ K for
1 $\mu$m wide connectors and   $T_{c3}=1.9$ K for the 80 nm-wide wire.  There
is a 4 nm thick wetting layer around Nb, which can be seen as a halo around
wires on the micrographs. For wires with small gaps $<100$ nm the wetting layer
fills the gap. The wetting layer is not attacked by the etching solution and
serves as an etch mask in the semiconductor wire definition. In the device
shown in Fig.~\ref{fs-GaAs} the gap is $\approx 40$nm. Yet these devices become
insulating when cooled to low temperatures. This test experiment allows us to
establish that (i) the thin Nb layer is not conducting and plays no role in the
electrical transport, (ii) the tunneling current for gaps $>20$ nm is
negligible, and (iii) in InSb JJs the current has to flow through the InSb
layer. \textbf{Thus the observed superconductivity in InSb JJs is due to the
proximity effect.}

\section{Temperature dependence of JJ}
\label{Ic-T-dependence}

\begin{figure}[h]
\def\ffile{fs-T-Tc}
\includegraphics[width=.6\textwidth]{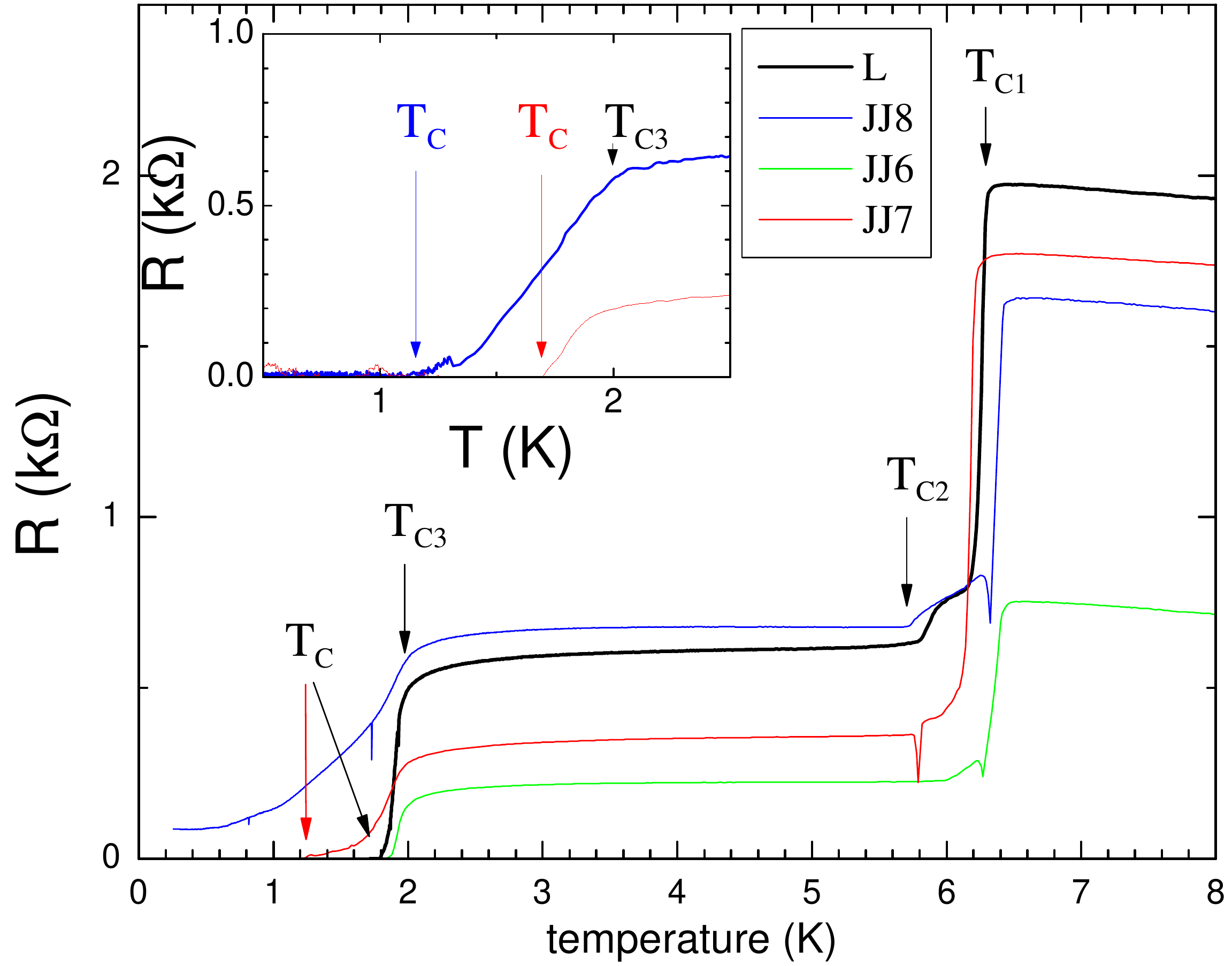}
\caption{Temperature dependence of resistance for the devices JJ6 (20 nm gap),
JJ7 (30 nm gap), JJ8 (40 nm gap) and a continuous 150 nm-wide line (L) is
measured in $^3$He system (main plot) and in a dilution refrigerator (inset).
The dilution refrigerator is properly shielded.}
\label{\ffile}
\end{figure}

Proximity-induced superconductivity has a reduced gap $\Delta$ compared to the
gap of the host superconductor, Eq.~(\ref{delta}). The lowest $T_{c}$
corresponds to this reduced proximity gap.

Temperature dependence of samples resistance is shown in Fig. ~\ref{fs-T-Tc}.
For a continuous line (L) we can identify three transition temperatures:
$T_{C1}=6.3$ K is for wide regions, $T_{C2}=5.7$ K is for 1 $\mu$m-wide wires
and $T_{C3}=1.9$ K is for the 150 nm-wide wire. Similar results are obtained
for Nb on GaAs, Section~\ref{GaAs}. Devices with a Josephson junctions (JJ6-8)
have the actual superconducting transition $0<T_C<T_{C3}$ for various devices.
Note that in a $^3$He refrigerator, where electrical noise is higher, we do not
observe the superconducting transition for the JJ8 device down to 250 mK, the
actual $T_C$ for this device is 1.17 K, see inset.

\section{Magnetic field dependence of critical current}

\begin{figure}[h]
\def\ffile{fs-Ic-B}
\includegraphics[width=.7\textwidth]{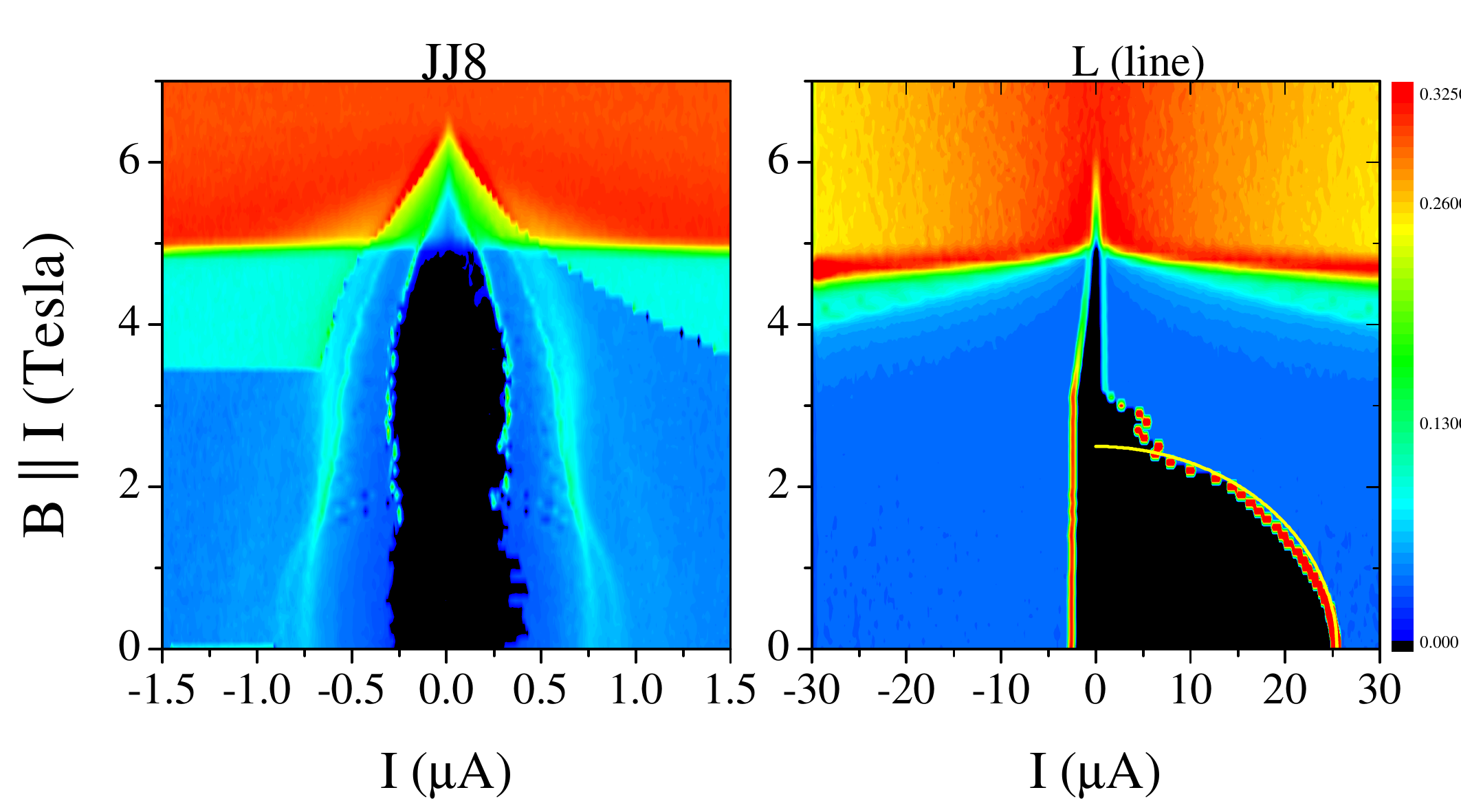}
\caption{Magnetic field dependence of differential resistance for device JJ8
and a 150 nm wide wire fabricated next to it (L). Black regions are the
superconducting states. Yellow line is $I_c/\sqrt{1-(B/B_1)^2}$, where
$B_1=2.5$ T is the first critical field.}
\label{\ffile}
\end{figure}

The requirement is that the proximity gap $\Delta>k_B T, h\omega_J$ in the
semiconductor material at the magnetic fields needed to satisfy the condition
$E_Z>E_F$. In our Nb wires superconductivity survives up to 5 Tesla, see
Fig.~\ref{fs-Ic-B}. In the low field region $I_c(B)\approx
I_c(0)/\sqrt{1-(B/B_1)^2}$ with $B_1\approx 2.5$ T. Resistivity of crystalline
Nb is 152 n$\Omega\cdot$m and a perfect $L=1.2\ \mu$m, $w=100$ nm and $t=40$ nm
wire should have resistance of $\approx 1\ \Omega$. Our wires have resistance
of $2-3$ k$\Omega$, which indicates a substantial degree of disorder. Neither
$\Delta$ nor $T_c$ are expected to be substantially affected by the disorder
\cite{Tinkham-book96}, consistent with the measured $T_c\approx 7$ K compared
to the $T_c=9.2$ K in pure crystalline Nb. Disordered Nb is a type-II
superconductor, and superconducting gap survives to much higher field
$B_2\approx 5$ T. The thickness of the film $t=40$ nm is of the order of the
coherence length in Nb (39 nm), thus flux capturing is possible even for the
in-plane field.

\section{Frequency dependence of Shapiro steps}

In Fig.~\ref{fs-f0a} we show $V(I)$ traces measured in the presence of rf field
with frequencies $f_0=2$, 3 and 4  GHz. The corresponding step heights are,
respectively, $\Delta V=hf_0/2e=4$, 6 and 8 $\mu$V.

\begin{figure}[h]
\def\ffile{fs-f0a}
\includegraphics[width=0.8\textwidth]{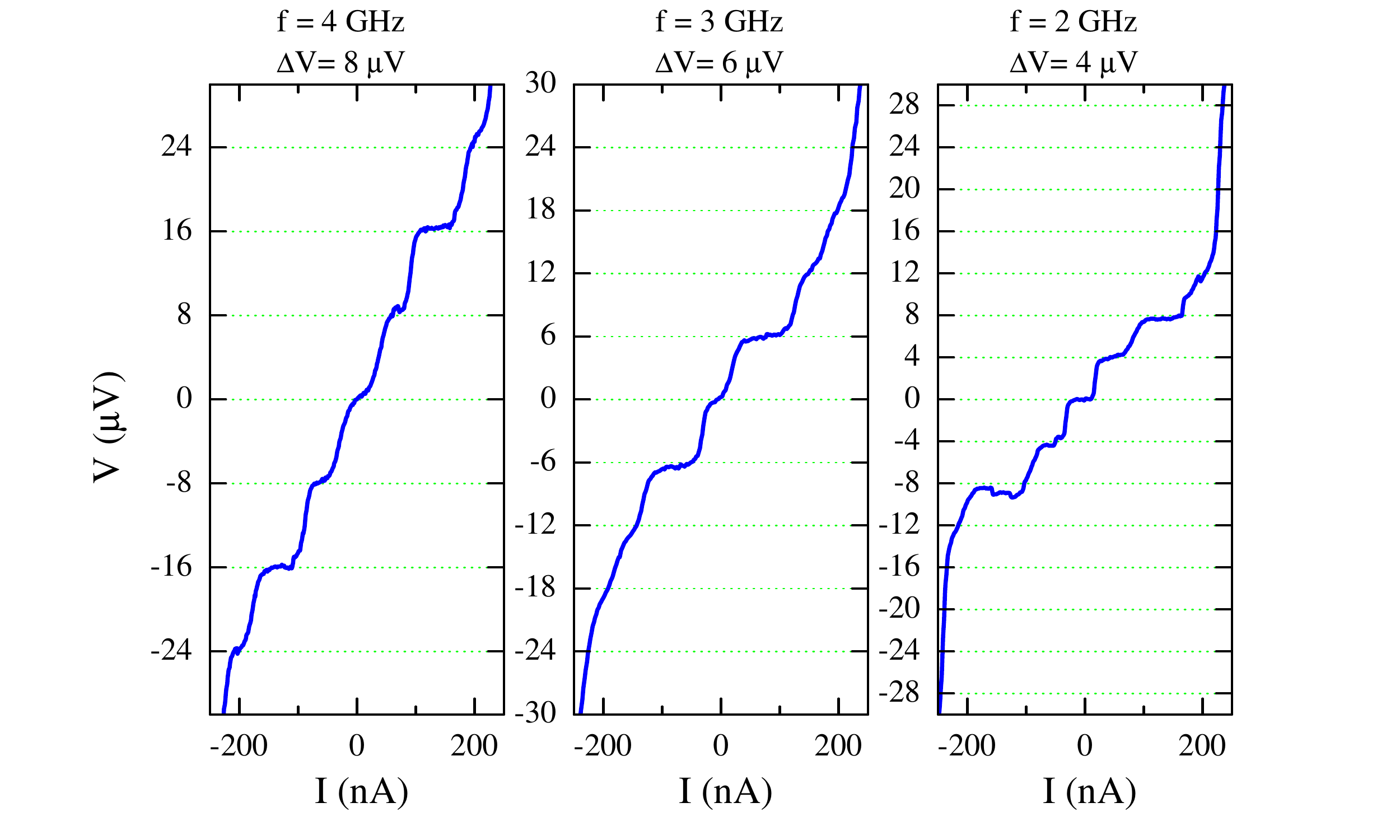}
\caption{Shapiro steps for rf frequencies  $f_0=2$, 3 and 4  GHz.}
\label{\ffile}
\end{figure}

\section{Analysis of the Josephson junction}

The wafers used in these experiments have substantial leakage between the Nb
film and a heavily doped substrate, and the substrate forms a shunting
resistance to every device. We can estimate the shunting resistance $R_s$ and
the normal resistance $R_N$ of the JJ8 as follows. Next to the JJ we fabricate
a 0.12 $\mu$m-wide wire with a measured resistance in the normal state of
$(R_w^{-1}+R_{sw}^{-1})^{-1}=0.6$ k$\Omega$. An identical wire on an insulating
GaAs substrate has a resistance of $R_w=2.5$ k$\Omega$, thus the shunting
resistance for the wire device $R_{sw}\approx 0.8$ k$\Omega$. The wire device
and the JJ devices have similar layouts and we assume that the shunting
resistances are similar, $R_s\approx R_{sw}$ (the leakage to the substrate is
primarily through contact pads of the same size). The measured resistance of
the JJ8 in the normal state is $(R_N^{-1}+R_{s}^{-1})^{-1}=0.65$ k$\Omega$,
thus we estimate $R_N\approx3$ k$\Omega$. The JJ8 device consists of two
nominally identical JJs in parallel, and the actual $R_N$ for the dominant
junction can be as high as $\approx6$ k$\Omega$. Thus, JJ8 has a few ($<5$)
subbands, which is also consistent with the number of subbands estimated from
the size quantization, see Section \ref{samples}.

For short weak links in the dirty regime ($L_{eff}\ll \xi$ and $l\ll L_{eff}$,
where $l$ is the mean free path, $\xi$ is the coherence length, and $L_{eff}$
is the effective length of the weak link), the product $I_c R_N[\mu
\mathrm{V}]\approx 320 T_c[\mathrm{K}]$ \cite{Likharev1979}. For the clean
regime ($L_{eff}\ll l, \xi$) the value should be almost twice as large, $I_c
R_N[\mu \mathrm{V}]\approx 480 T_c[\mathrm{K}]$. From the above estimates the
product $I_c R_N\approx1$ mV, while $T_c=1.17$ K. Thus, JJ8 is in the clean
regime, which is favorable for the observation of Majorana fermions.

\section{Analysis of Shapiro steps as a function of rf power}
\label{rf-power}

\begin{figure}
\def\ffile{fs-dVdI-dI}
\includegraphics[width=1\textwidth]{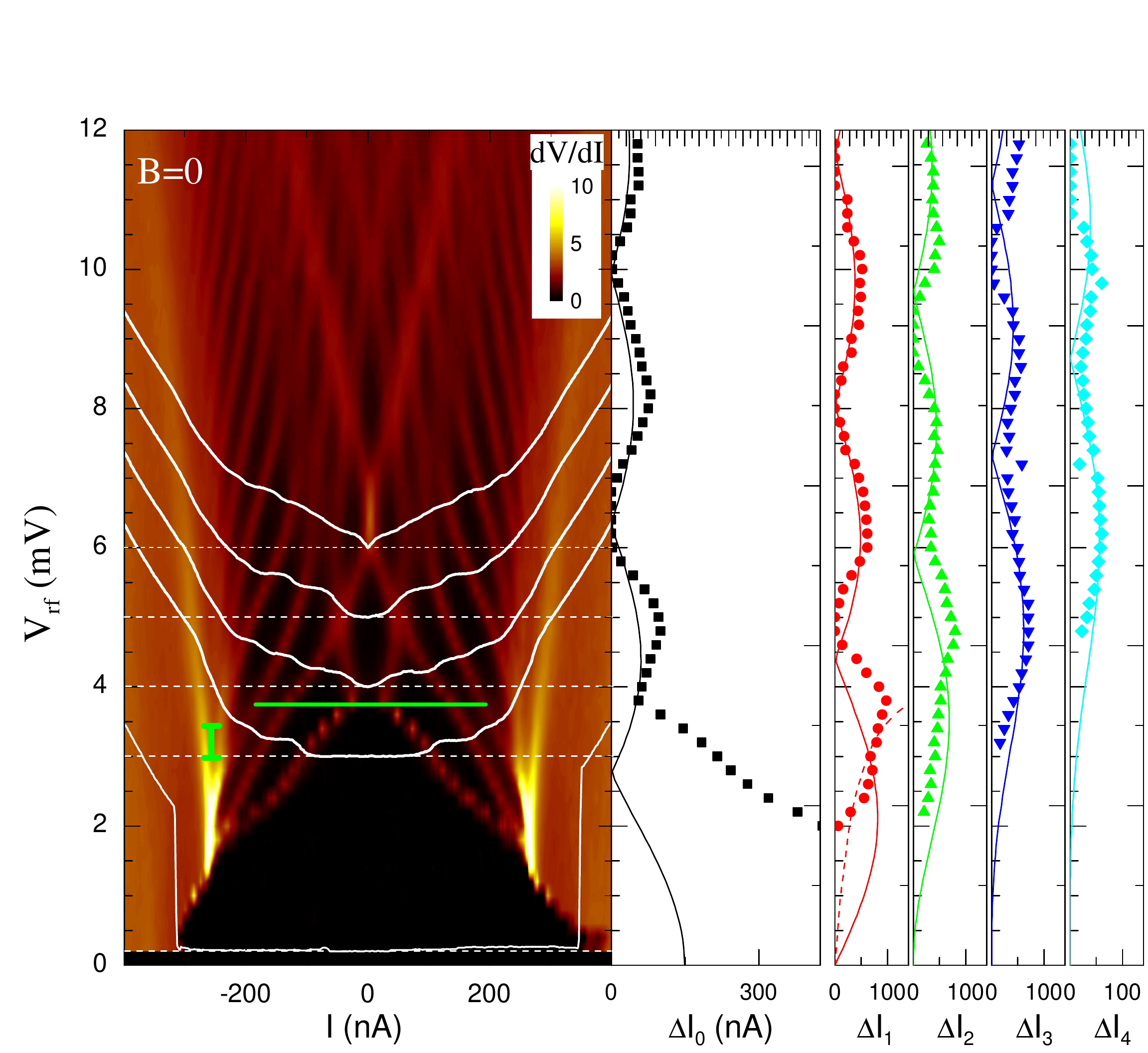}
\caption{Left: differential resistance  $dV/dI$ of the JJ8 is plotted as a
function of the rf amplitude $V_{rf}$ and dc current $I$. The rf frequency is
$f_0=3$ GHz. The data is measured with low frequency (17 Hz) ac excitation
$I_{ac}=2$ nA at $T=20$ mK and $B=0$.  $|V(I)|$ characteristics at
$V_{rf}=0.2$, 3, 4, 5 \& 6 mV are shown as white lines, their zero is shifted
vertically and is marked by dashed white lines. The short vertical bar is a 12
$\mu$V scale. Right: $n$-th step width $\Delta I_n$ is extracted from the
$dV/dI$ data and plotted as a function of $V_{rf}$ for $n=0-4$. Solid lines are
Bessel functions $A|J_n(\beta V_{rf})|$ with $A=150$ nA and $\beta=0.89$
mV$^{-1}$. A red dashed line for $n=1$ is $\propto V_{rf}/\bar{I}$.}
\label{\ffile}
\end{figure}

In Fig.~\ref{fs-dVdI-dI} we plot  differential resistance $dV/dI(I,V_{rf})$ at
$B=0$ for the JJ8 device. Shapiro steps with $\Delta V=6$ $\mu$V are clearly
seen at $V_{rf}>2$ mV and correspond to the black regions on the $dV/dI$
contour plot. At high $V_{rf}>4$ mV evolution of the width of the Shapiro steps
$\Delta I_n$ follows the Bessel function pattern as a function of power,
$\Delta I_n=A|J_n(2e v_{rf}/hf_0)|$, where $v_{rf}$ is the rf amplitude at the
junction and $f_0$ is the rf frequency. We can estimate the rf power
attenuation from the fitting of $\Delta I_0$ with $J_0$ above $V_{rf}>4$ mV,
$v_{rf}=5.4\cdot 10^{-3} V_{rf}$ at $f_0=3$ GHz and $v_{rf}=1.7\cdot 10^{-3}
V_{rf}$ at $f_0=4$ GHz. Here $V_{rf}$ is the rf amplitude at the top of the
fridge.

In the $dV/dI(I,V_{rf})$ plot we can identify two regions with different
$\Delta I_0(V_{rf})$ dependencies, separated schematically by a green line at
$V_{rf}^{crit}\approx 3.7$ mV. For the JJ8 device $V_c=I_R R_N\approx 65\
\mu\mathrm{eV}\gg hf_0/2e=6 \mu\mathrm{eV}$, see Fig.~2 of the manuscript, and
for $V_{rf}<V_{rf}^{crit}$ the junction is in the small microwave signal regime
\cite{LikharevJJbook}. The linear response of a JJ has a singularity at
$\omega=\pm\omega_J$, and the first Shapiro step appears due to phase locking
of the external frequency and the Josephson oscillations. The JJ8 is in the
intermediate damping (resistively shunted) regime, and the $V(I)$
characteristic is expected to become non-hysteretic in the vicinity of the
first step. Indeed, we observe no hysteresis for $V_{rf}>1.8$ mV. \textbf{While
nonlinear effects can be present at high $I$ and $V_{rf}$ we want to stress
that the first step at the onset of the normal state is due to phase locking
between the external and the Josephson frequencies, $\omega=\pm\omega_J$.}

The width of the first step in the low rf power regime is expected to be
$\Delta I_1=2I_c I_{\omega}/\bar{I}$; see dashed red line on $\Delta
I_1(V_{rf})$ plot. At high power the width of the $n$-th step is expected to
follow the $n$-th Bessel function. Indeed the step widths follow the expected
pattern, but some deviations are expected due the sample being in a crossover
regime between the low and the high rf power regimes.

\section{Magnetic field dependence of the Shapiro steps}

\begin{figure}[p]
\def\ffile{fs-dR-Vrf}
\includegraphics[width=1.0\textwidth]{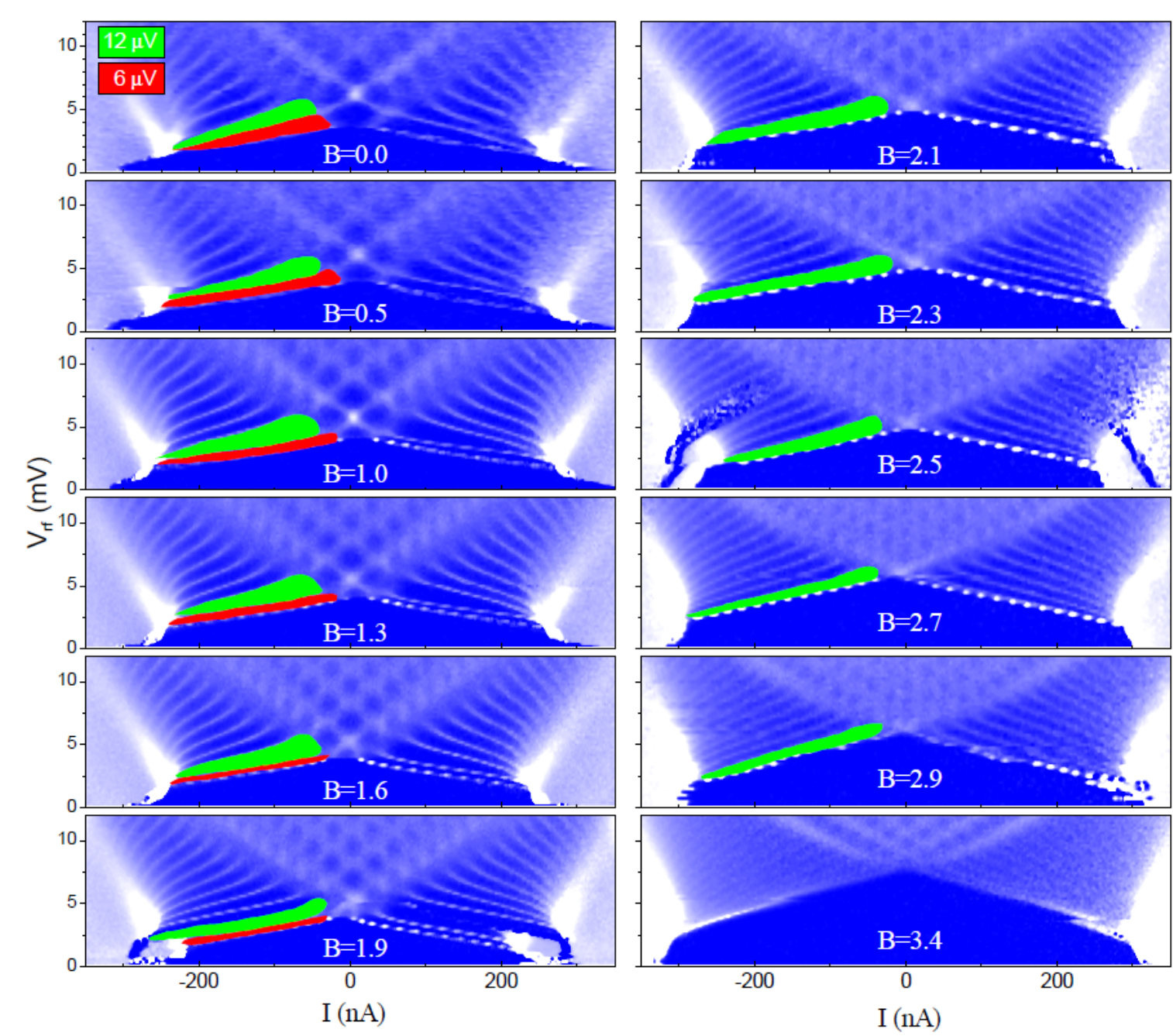}
\caption{Evolution of Shapiro steps in JJ8 with magnetic field. The data is
measured with $I_{ac}=2$ nA at $T=20$ mK. The steps at 6 $\mu$V and 12 $\mu$V
are outlined with red and green, respectively. We use $V(I)$ data to identify
the step height.  Note that the  6 $\mu$V step disappears for $B>2$ Tesla.}
\label{\ffile}
\end{figure}

A detailed field dependence of step evolution is given in Fig.~\ref{fs-dR-Vrf},
and the analysis of step width as a function of rf power is discussed in
Section \ref{rf-power}.

\bibliographystyle{naturemag}
\bibliography{rohi}

\end{document}